\documentclass[conference]{IEEEtran}
\IEEEoverridecommandlockouts

\usepackage{subcaption}
\usepackage{cite}
\usepackage{amsmath,amssymb,amsfonts}
\usepackage{algpseudocode}
\usepackage{graphicx}
\usepackage{textcomp}
\usepackage{svg}
\usepackage{multirow}
\usepackage{adjustbox}
\def\BibTeX{{\rm B\kern-.05em{\sc i\kern-.025em b}\kern-.08em
    T\kern-.1667em\lower.7ex\hbox{E}\kern-.125emX}}

\usepackage{colortbl}
\usepackage{algorithm}
\usepackage{array}
\usepackage[caption=false,font=normalsize,labelfont=sf,textfont=sf]{subfig}
\usepackage{stfloats}
\usepackage{url}
\usepackage{verbatim}
\begin{document}

\title{FastRAG: Retrieval Augmented Generation for Semi-structured Data}

\author{
\IEEEauthorblockN{Amar Abane}
\IEEEauthorblockA{\textit{NIST} \\
Gaithersburg, USA \\
amar.abane@nist.gov}
\and
\IEEEauthorblockN{Anis Bekri}
\IEEEauthorblockA{\textit{NIST} \\
Gaithersburg, USA \\
anis.bekri@nist.gov}
\and
\IEEEauthorblockN{Abdella Battou}
\IEEEauthorblockA{\textit{NIST} \\
Gaithersburg, USA \\
abdella.battou@nist.gov}
\and
\IEEEauthorblockN{Saddek Bensalem}
\IEEEauthorblockA{\textit{Université Grenoble Alpes} \\
France \\
Saddek.Bensalem@univ-grenoble-alpes.fr}
}

\maketitle

\begin{abstract}
Recent advances in Large Language Models (LLM) and Retrieval-Augmented Generation (RAG) techniques have improved data processing in network management. However, existing RAG methods like VectorRAG and GraphRAG struggle with the complexity and implicit nature of semi-structured technical data, leading to inefficiencies in time, cost, and retrieval.
This paper introduces FastRAG, a novel RAG approach for semi-structured data. FastRAG proposes chunk sampling, schema learning, and script learning to extract and structure data without submitting entire data sources to the LLM. It integrates text search with knowledge graph (KG) querying to improve accuracy. The evaluation results demonstrate that FastRAG provides accurate question answering while improving up to $90\%$ in time and $85\%$ in cost compared to GraphRAG.
\end{abstract}


\section{Introduction}
Efficiently processing and understanding semi-structured data is crucial to improving network management tasks. As networks expand, the diversity of platforms and the sheer volume of data introduce significant challenges, including operational inefficiencies and delays in issue resolution. Traditional tools for processing network data \cite{min2017Deeplog, batfish2023, dna} provide some utility, yet they fail to extract and leverage the full range of information embedded in semi-structured data formats such as logs and configurations. These limitations often hinder the ability to correlate, analyze, and act on insights, particularly in complex environments that involve heterogeneous devices and platforms.

Disaggregated implementations of network services further exacerbate the task of correlating semi-structured data from different vendors. This fragmentation not only complicates the parsing and normalization of data but also makes it challenging to derive actionable insights across varying formats. Although natural language processing (NLP) and machine learning (ML) algorithms have been developed to address these challenges \cite{huo2023SemparserLog, birkner2022Config2spec}, they often lack the flexibility needed to handle the diversity of data types and formats encountered in real-world network environments.

The emergence of Large Language Models (LLMs) has introduced new possibilities for processing and understanding natural language, showcasing their potential in improving network management tools by automating complex analytical tasks. However, using pre-trained models to generate domain-specific and coherent responses remains challenging. Techniques like Retrieval-Augmented Generation (RAG) \cite{gao2024RagSurvey} have been introduced to enhance LLMs by integrating retrieval methods, with VectorRAG providing context retrieval from textual documents based on semantic similarity \cite{lewis2020RagNlp}. Although effective in many scenarios, these traditional RAG systems have limitations, such as loss of critical contextual information due to uniform fragmentation of documents \cite{sarmah2024HybridRag}.

GraphRAG \cite{edge2024GraphRag}, a more advanced approach, addresses some of these limitations by leveraging knowledge graphs (KG) to organize information extracted from source documents. Despite its advantages, GraphRAG struggles with queries for accurate information or queries lacking explicit entities. HybridRAG \cite{sarmah2024HybridRag} further refines these methods by combining the strengths of VectorRAG and GraphRAG, improving the precision in retrieving relevant information for LLM.

Despite these advancements, existing RAG techniques do not meet the unique challenges posed by network data. The reliance on embedding vectors for context retrieval often underperforms with semi-structured technical data, where implicit information is frequently encoded in domain-specific keywords. Furthermore, current RAG systems rely heavily on LLMs to extract structured information from source documents by processing them chunk by chunk, leading to increased computational costs and inefficiencies, critical drawbacks when dealing with the large-scale, high-frequency data generated in modern networks.

To address these issues, we propose FastRAG, a novel RAG approach designed to process large volumes of semi-structured data in a cost-effective way. FastRAG addresses the specific needs of network management by offering a system optimized to extract actionable insights from diverse data sources while minimizing latency and cost. 
The system introduces innovative methods, including schema learning and script learning for structured data processing, an algorithm to select representative chunks for learning prompts, and a hybrid retrieval approach that integrates text search with graph querying using a query language.
These methods enable FastRAG to deliver improved accuracy and efficiency, making it suitable for network management tasks such as troubleshooting, configuration auditing, and network analysis.

The remainder of this paper is organized as follows. Section \ref{rw} reviews related work and discusses the motivation behind the development of FastRAG. Section \ref{design} presents the detailed design of the FastRAG system, including its key components and techniques. The evaluation of FastRAG is discussed in Section \ref{evaluation}. Finally, Section \ref{conclusion} outlines the system's limitations and concludes the paper.

\section{Related Work and Motivation}\label{rw}
The integration of LLMs into network management has opened new possibilities, particularly in automating network configuration tasks. Recent studies have explored the potential of LLMs to generate network configurations \cite{jeong2024switchLlm,rajdeep2023LlmRouterConfig,wang2023NetworkConfigHuman}, detect anomalies \cite{le2023LogParsingPrompt}, and convert textual descriptions into formal specifications \cite{donadel2024LlmUnderstandComputerNetworks, ifland2024GenetMultimodalCopilot}. 
These contributions represent the most frequent applications of LLMs in network management, aiming to reduce manual labor and error risks.
The work proposed in this paper is complementary to these approaches, as they may rely on RAG systems to retrieve network-specific information. 

The application of RAG within network management has been less explored. RAGLog \cite{pan2023RagLog} introduces a system for detecting anomalies using a combination of vector database and LLM, supporting raw log data and various log sources without extensive pre-processing. Although promising for its adaptability, RAGLog faces resource consumption and latency challenges, which could limit its scalability in larger network data.
Telco-RAG \cite{bornea2024TelcoRag} targets the processing of 3GPP documents. A query enhancement pipeline demonstrates improved accuracy when handling complex questions related to the telecommunication domain. 

More generally, VectorRAG, the predominant approach in RAG systems, computes vector embeddings for fixed-size chunks and uses semantic similarity to retrieve information relevant to the input query. However, this method struggles with the nature of network data, where domain-specific keywords have different meanings in their context. 
GraphRAG \cite{edge2024GraphRag} is a more advanced approach that integrates KGs with text chunking and embedding. The LLM extracts entities and relationships and summarizes communities detected in the KG, providing a more structured retrieval (also using semantic similarity). 
Although GraphRAG is efficient in summarizing and reporting tasks, it falls short in delivering valuable answers when the query includes specific values, such as names or types, that need to be matched precisely in the retrieval. 
For example, to the typical network management question \textit{"What is the IP address of interface Ethernet0 on router r1?"} GraphRAG would either not be able to answer the question, or provide a summary of several pieces from configuration files that refer to the mentioned interface.
HybridRAG \cite{sarmah2024HybridRag} combines GraphRAG and VectorRAG. Although this hybrid approach improves the accuracy and contextual relevance of responses, it inherits the limitations of its base approaches. 
Another approach uses entity-relationship extraction to create a KG capturing the important information of the source data. The KG is then searched as a database by translating the textual query into a database query. Although this allows for accurate retrieval, it can be impractical given the imperfect parsing obtained with LLMs \cite{llm_accuracy}.

In contrast, FastRAG extracts entities and their properties from the source documents while mapping each entity to specific lines of the original text. This method enables accurate retrieval using a KG while allowing text search to handle vague queries and compensate for imperfect entity extraction. Unlike semantic similarity, the text search matches text based on the exact wording or structure. KG and text searches are performed using queries on a single KG implemented in a graph database. 
Unlike other methods, FastRAG avoids costly processing of all source data through LLMs by generating JSON schemas and Python code for data structuring and parsing. 
While demonstrated on logs and configuration data, this approach can be applied to other semi-structured data, such as playbooks and alarms. 
To our knowledge, this is the first RAG system to rely solely on code generation for data processing.

\section{FastRAG Design}\label{design}
We designed a hybrid approach for both information extraction and retrieval. 
Instead of using indiscriminate chunking or a user-defined schema, our method converts the data into a simple, automatically generated JSON structure while maintaining a link to the original data in a KG \cite{khorashadizadeh2024RTrendsLlmKg}. Our RAG also relies on LLMs to interact with the KG, utilizing query generation to retrieve relevant information.

Given the extensive and continuously updating nature of network data, minimizing processing time and cost is essential. To achieve this, we leverage prompt engineering findings that indicate that LLMs are more effective at generating code rather than directly extracting information into structured formats. Therefore, we introduce \textbf{schema learning} and \textbf{script learning}, where we \textbf{sample} the source data and iteratively prompt the LLM to: (i) generate JSON schemas to structure the information, and (ii) generate Python scripts to parse the source data and extract entities.

\begin{figure}[!h]
\centering
\includegraphics[width=\linewidth]{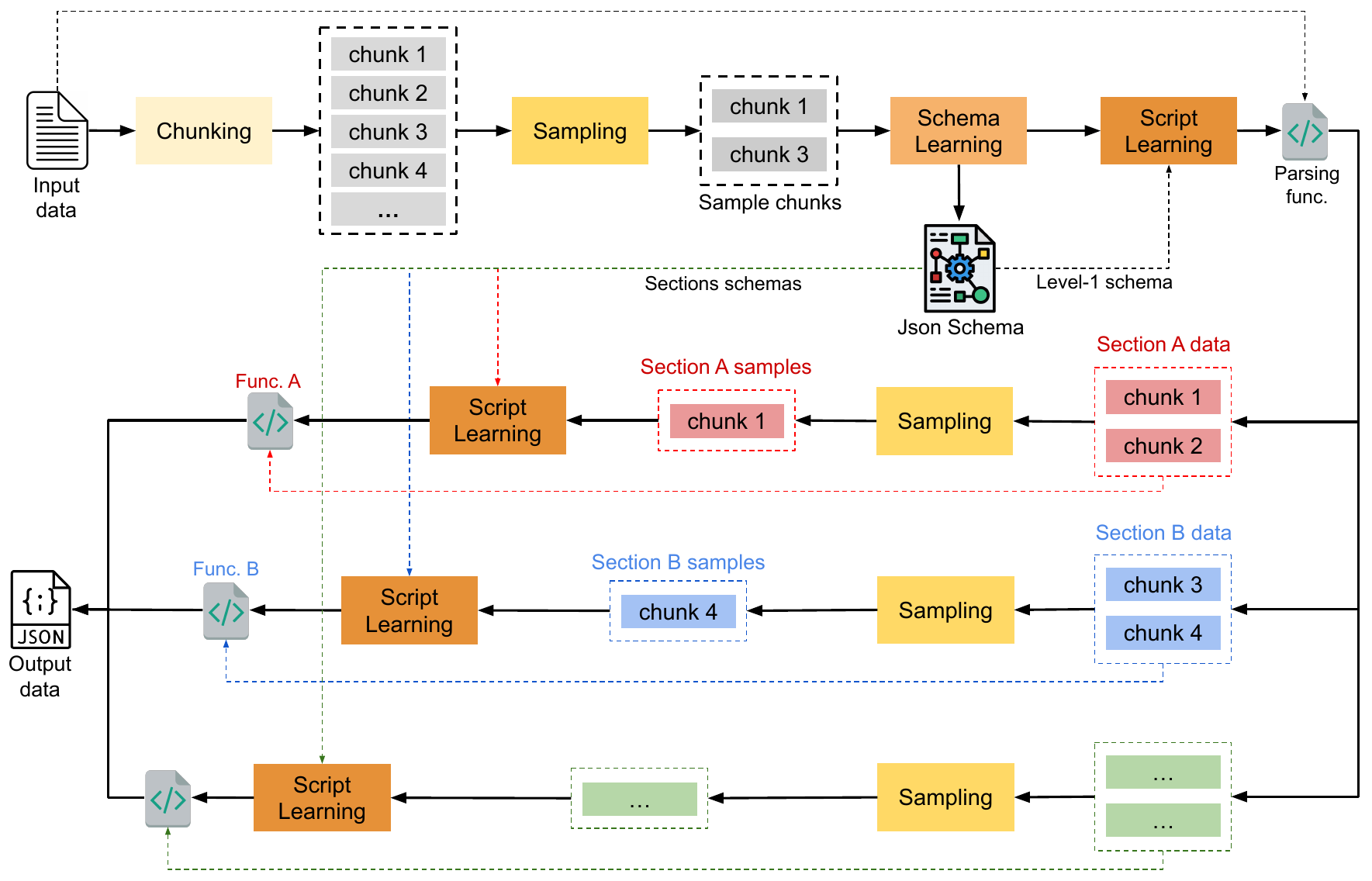}
\caption{FastRAG data processing architecture}
\label{fig:architecture}
\end{figure}

Figure \ref{fig:architecture} shows the FastRAG data processing pipeline, illustrating chunk sampling, schema learning, and script learning. Each of these methods is described below, followed by a description of the information retrieval process.

\subsection{Chunk sampling}
To address LLM input size constraints, data is usually partitioned into uniformly sized, overlapping chunks and submitted via batched or sequential queries. However, this approach incurs significant latency and computational cost. To address this, we exploit the repetitive formatting present in semi-structured data, selecting a representative set of chunks, or sample chunks, that effectively covers the syntax of the source data \cite{le2023LogParsingPrompt}.
The chunk sampling process involves two key procedures: keyword extraction and chunk selection.

\subsubsection{Keyword extraction}
We draw inspiration from NLP techniques to identify the most meaningful terms in the source data \cite{huo2023SemparserLog}. The process begins with the preprocessing of the text, where the punctuation is removed, and the content is tokenized into individual words. Given that English is the predominant language used in most semi-structured network data, it is adopted for this process. 
After preprocessing, the text is split into lines, and a matrix representation is created based on word frequency, where each row corresponds to a line, and each column to a unique term from the entire text. The matrix, which reflects the frequency of each term in each line, serves as the input for the K-means clustering algorithm. 
This algorithm groups the lines into $n_c$ groups based on the similarity of their frequency patterns. The $n_t$ terms closest to the centroids are selected within each group as keywords. The output is a set of keywords that encapsulate the primary terms present in the text corpus.

\subsubsection{Samples selection}
We developed an algorithm that selects the smallest set of chunks that contains all keywords. However, merely adding chunks until all keywords are covered does not guarantee the selection of the (near) minimum number of sample chunks. To address this, the algorithm selects chunks based on their entropy, providing an estimate of the diversity introduced by each chunk. 

The procedure (see Algorithm \ref{algorithm}) begins with preprocessing of the chunks (as in the previous step) where only the extracted keywords are retained. The preprocessed chunks are then used to compute Term Frequency-Inverse Document Frequency (TF-IDF) vectors, representing the importance of terms within each chunk. The Shannon entropy for each chunk is then calculated to gauge the informational diversity to which it contributes. The algorithm iteratively selects a subset of chunks to maximize term coverage, focusing on those that introduce new terms and the most diversity, until the desired coverage threshold - fixed at \textit{1} in all cases - is reached.

\begin{algorithm}[!th]
\caption{Samples Selection}
\begin{algorithmic}
\State \textbf{Input:} $chunks, threshold, keywords$
\State $tokenized \gets \text{tokenize/filter chunks w/ } keywords$
\State $tfidf \gets \text{TF-IDF vectors of } tokenized$
\State $entropies \gets \text{entropy for each chunk in } tokenized$
\State $n\_chunks, n\_terms \gets \text{from } tfidf$
\State $selected, covered \gets \emptyset, \emptyset$
\While{$\frac{|covered|}{n\_terms} < threshold$}
    \State $gains \gets \emptyset$
    \For{each $i \in [1, n\_chunks]$}
        \If{$i \notin selected$}
            \State $nw\_trms \gets \text{terms in $tokenized[i]$ not in } covered$
            \State $gain \gets |nw\_trms| \times entropies[i]$
            \State $gains \gets gains \cup (gain, i)$
        \EndIf
    \EndFor
    \If{$gains \neq \emptyset$}
        \State $best \gets \text{chunk w/ max gain from } gains$
        \State $selected \gets selected \cup best$
        \State $covered \gets covered \cup \text{terms in } tokenized[best]$
    \Else
        \State \textbf{break}
    \EndIf
\EndWhile
\State $output \gets \text{chunks corresponding to } selected$
\State \textbf{Output: } $output$
\end{algorithmic}
\label{algorithm}
\end{algorithm}

The number of sample chunks selected is directly influenced by the number of keywords extracted, which can be adjusted by varying the parameters $n_c$ (number of clusters) and $n_t$ (number of terms).
This process is discussed in more detail in Section \ref{evaluation}.

\subsection{Schema learning}
Building on the selected sample chunks, we develop a prompt strategy that guides the LLM in identifying entity types and their properties, focusing on schema extraction rather than specific entities.
The process (Figure \ref{fig:llm_learning}) begins with the first sample chunk, submitted to the LLM to identify and structure the types of entities and their properties in a JSON schema. The JSON schema is refined in subsequent prompts by submitting new chunks to the LLM. To avoid complexity, the LLM is instructed to limit high levels of nesting and refrain from creating objects with an excessive number of properties.

After each prompt, the JSON schema is validated. If the schema contains errors, the LLM is asked to correct them based on the error message provided. Each prompt call can be retried up to $N=4$ times.

\begin{figure}[!h]
\centering
\includegraphics[width=\linewidth]{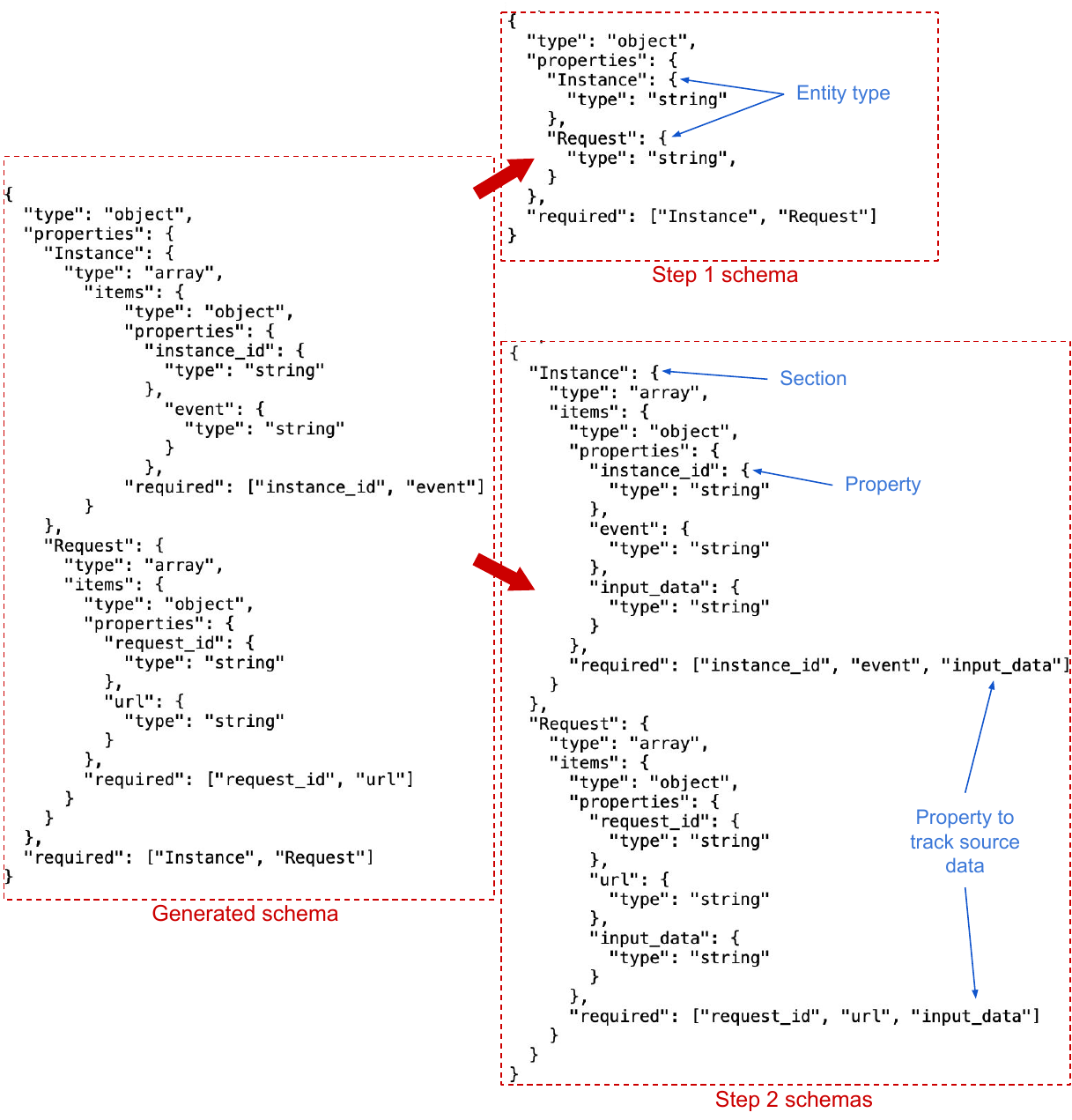}
\caption{Example of processing of generated schema}
\label{fig:schema}
\end{figure}

Two types of objects are extracted from the finalized schema, as illustrated in Figure \ref{fig:schema}. The first is the \textit{Step 1} schema, which is a JSON schema that includes only the level-1 entity types -referred to as sections. Each entity type includes a description\footnote{Not shown in the Figure} and a string property containing the source data lines that define the objects that belong to that section. 
The second, \textit{Step 2} schema, is a map where each section is associated with an array where objects correspond to the entire schema of that section.
This division into Step 1 and Step 2 schemas allows for a more structured approach to data extraction, as discussed below.

\subsection{Script learning}
\begin{figure}[!h]
\centering
\includegraphics[width=.5\linewidth]{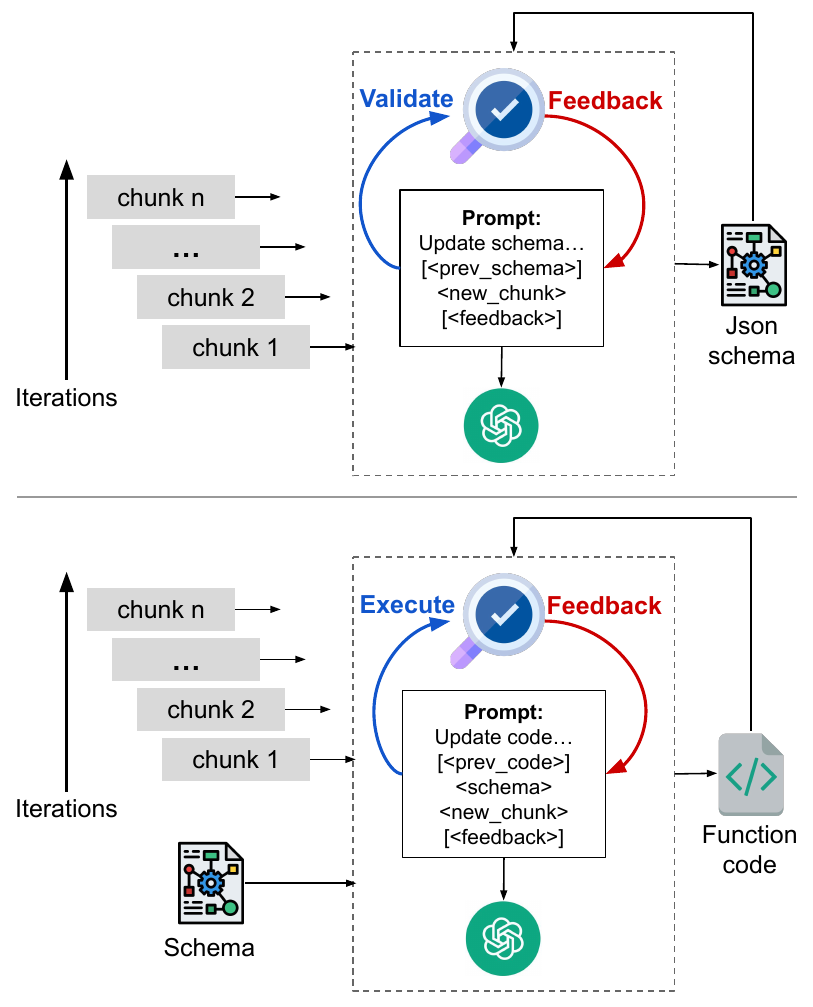}
\caption{Schema learning (top) and script learning (bottom)}
\label{fig:llm_learning}
\end{figure}

Similar to the approach used in schema learning, the script learning prompt strategy (Figure \ref{fig:llm_learning}) begins by instructing the LLM to generate parsing functions based on the first sample chunk. The LLM is then prompted iteratively to refine the previous function code by submitting new sample chunks. 
To ensure both the syntactic and functional correctness of the generated code, the prompt includes a verification step, where the code is executed with the sample chunk used to generate it. If the code contains errors, the LLM is asked to correct them based on the feedback message. Similarly, each prompt call can be retired up to N times.

In Step 1, the prompt instructs the LLM to map each line in the source data to its corresponding section according to the schema. Once the function is generated, it is used to parse the entire source data and divide it into sections. 
After that, each section is further split into fixed-size chunks, and the sampling process is repeated for each section. At this stage, each section has its schema (contained in the schema object in Step 2) and a set of sample chunks, enabling it to be processed independently, following the same method as in Step 1.

In Step 2, script learning is utilized to generate a specific parsing function for each section. Once these functions are generated, the data within each section is processed using the corresponding parsing function.
Processing sections independently enables the LLM to concentrate more effectively on each entity type and also allows for a more targeted refinement of the sample chunks by re-executing the sampling process for each section.

\subsection{Information Retrieval}
The extraction process results in a JSON object where each entity is represented, and within this object, an \textit{input\_data} property contains the lines from the source data that define the entity. This JSON structure serves as the foundation for the creation of KGs and the retrieval of information.

\subsubsection{KG Creation}
Each entity identified in the JSON object is inserted as a node within the KG, the entity type serving as the node's label. Simple-type properties of the entity are directly assigned as properties of the node, while properties that are themselves objects are inserted as child nodes, connected to their parent node. Furthermore, for each line within the \textit{input\_data} property of each entity, a corresponding node is created and linked to the parent entity node. 
These input data nodes are used for text searches that utilize NLP methods \cite{apacheLucene}.

Several retrieval strategies have been defined and tested to interact with this KG, as discussed below.

\subsubsection{KG Querying (Graph)}
A prompt is used to provide the LLM with the KG schema. The LLM is then instructed to generate a syntactically correct query statement that can answer the user input. The generated query is executed, and the results are interpreted by the LLM to generate the final answer.

\subsubsection{Text Search (Text)}
The LLM is given examples of text search features (e.g., from the documentation) and instructed to generate a query that exclusively utilizes text search features to respond to user input. The query is then executed, and the results are interpreted by the LLM.

\subsubsection{Combined Querying (Combined)}
This strategy involves running both the KG query and the text search prompts in parallel for the input of the user. The raw results of each method are then provided to the LLM to synthesize the final answer.

\subsubsection{Hybrid Querying (Hybrid)}
This strategy merges the features of KG querying and text search on the same prompt. The prompt provided to the LLM includes both the KG schema and text search examples and instructs the LLM to generate a query that can use any relevant feature(s) to answer the user input.

\section{Implementation and Evaluation}\label{evaluation}
To implement FastRAG, we used the OpenAI GPT-4o model as LLM, Neo4j as the graph database to store and query the KG using Cypher and the Langchain library.

The evaluation used two different data sources that represent typical network data: logs and configurations. 
For the logs, we used a single file containing 2000 lines of OpenStack logs \cite{zhu2023Loghub}, resulting in 1307 chunks of $\sim1000$ tokens each. 
For the configurations, we used 13 Cisco device configuration files, totaling 2100 lines, and divided them into 19 chunks of $\sim1000$ tokens each.

The sizes were intentionally kept small due to their repetitive syntax and semantics. The heuristic discussed below for determining optimal samples generalizes to larger datasets, making larger size unnecessary for performance evaluation. This approach also allowed for a focused understanding of the data's content to evaluate question answering.

\subsection{Information Extraction}
\subsubsection{Step 1 schema and script learning}
Determining the optimal number of sample chunks for each dataset is crucial, as it directly affects the quality of schema and script learning. If too few samples are used, the system may overlook important entity types, leading to incomplete parsing. However, too many samples can result in overly complex schemas with unnecessary entity types. Furthermore, we observed that the optimal sample size for Step 1 does not necessarily translate to Step 2, as the keywords extracted in Step 1 might not fully align with the relevant keywords for each section in Step 2. Consequently, it is necessary to find the best number of sample chunks for each step independently.

The size of the samples is a hyperparameter that requires experimentation with different sizes by varying $n_c$ and $n_t$ to identify the sample size that produces the best extraction of information. 
To streamline this process, we defined a \textit{coverage} metric. This metric measures the ratio of lines in the source documents that are included in the \textit{input\_data} of all extracted objects.

\begin{figure}[!t]
\centering
\begin{subfigure}[b]{0.48\linewidth}
    \centering
    \includegraphics[width=\linewidth]{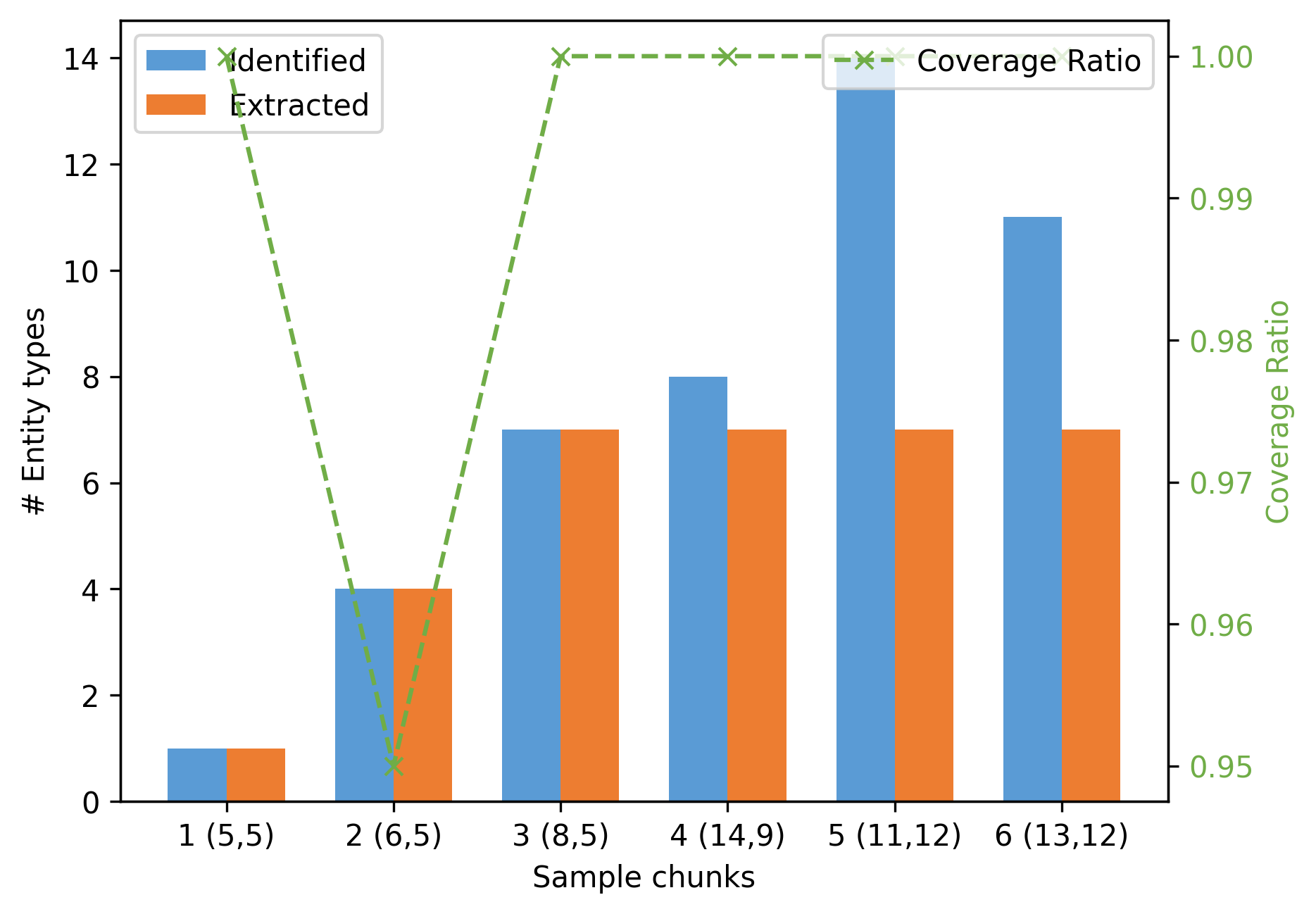}
    \caption{Logs}
    \label{fig:step1_logs}
\end{subfigure}
\hfill
\begin{subfigure}[b]{0.48\linewidth}
    \centering
    \includegraphics[width=\linewidth]{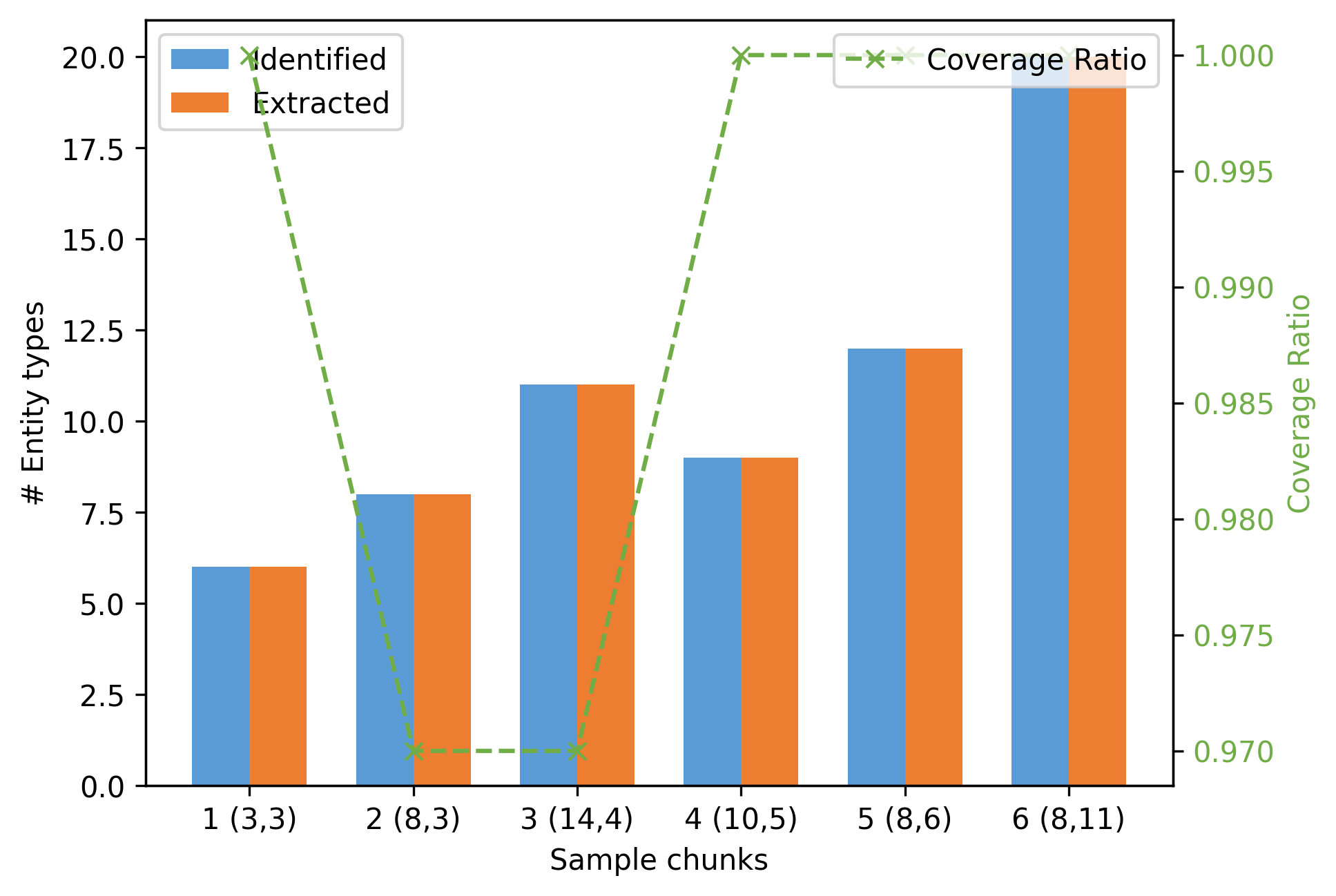}
    \caption{Configurations}
    \label{fig:step1_configurations}
\end{subfigure}
\caption{Effect of sample size on extraction (Step 1)}
\label{fig:step1_sample_size}
\end{figure}

To determine the optimal number of chunks for Step 1, we varied the sample size from 1 to M by adjusting $n_c$ and $n_t$ to achieve the desired number. We then ran the schema extraction process, generated the corresponding parsing function, and calculated the coverage ratio by executing the function on the source data.
Figure \ref{fig:step1_sample_size} reports the results of this procedure. The coverage ratio reaches 1 with just one sample chunk; however, the schema produced is overly simplistic, identifying only one entity type. Therefore, it is also important to consider the completeness and relevance of the extracted schema by examining the number of identified entity types. The goal is to select the most comprehensive schema that leads (near) to $100\%$ coverage ratio.
Another observation is that the LLM tends to add more entity types when more chunks are used, possibly due to the instruction to avoid too many properties in a single entity type. However, we found that overlapping or superfluous entity types are not a significant concern. During script learning, the LLM often fails to find efficient ways to parse these extra types, resulting in empty sections that are ignored in Step 2. This occurs because the LLM prioritizes achieving a 100\% coverage ratio, sometimes at the expense of extracting all entity types.
To highlight this, Figure \ref{fig:step1_sample_size} differentiates between the number of entity types identified and the number of entity types extracted with the generated parsing function. Based on these observations, we determined the optimal number of sample chunks for each dataset: 4 chunks out of 1307 for the logs dataset and 4 chunks out of 19 for the configuration dataset.

In addition to aiding in the selection of the best sample size, the coverage ratio also serves as an empirical metric to evaluate the quality of extraction of our RAG system.

\begin{figure}[!t]
\centering
\begin{subfigure}[b]{0.48\linewidth}
    \centering
    \includegraphics[width=\linewidth]{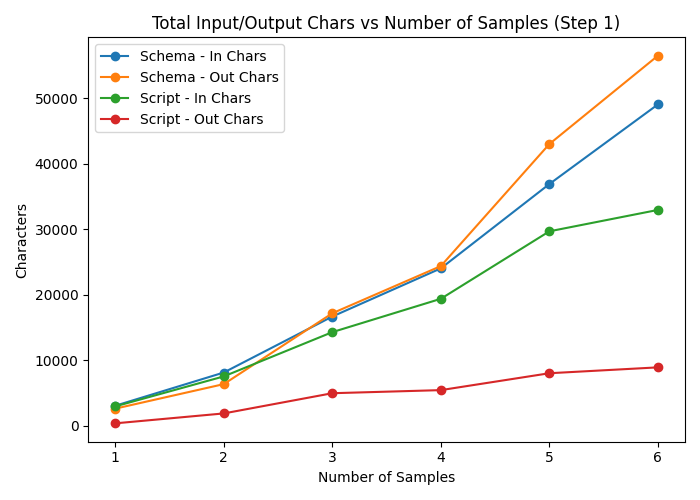}
    \caption{Logs}
    \label{fig:chars_logs} 
\end{subfigure}
\hfill
\begin{subfigure}[b]{0.48\linewidth}
    \centering
    \includegraphics[width=\linewidth]{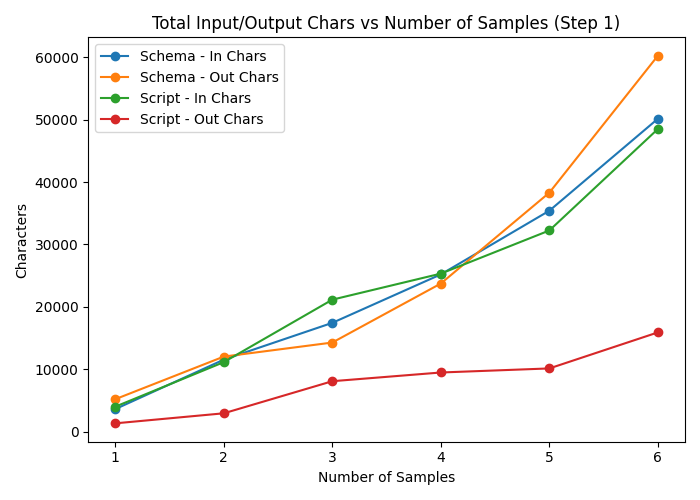}
    \caption{Configurations}
    \label{fig:chars_configs} 
\end{subfigure}
\caption{Effect of sample size on input and output size (Step 1)}
\label{fig:step1_chars}
\end{figure}

To assess the efficiency of using the LLM in our method, we measured the volume of input and output characters exchanged with the LLM in both the schema and script learning prompts, shown in Figure \ref{fig:step1_chars}. The input character count for both prompts is similar, particularly in the first half of the chart. However, output characters, which weigh more in the model's pricing and are limited to 4000 tokens per output, show a different trend. Script learning tends to use fewer output characters, whereas schema learning's output increases significantly with larger sample sizes. However, with the sample size fixed to $4$ for both datasets, the output character count for schema learning remains within a reasonable range, indicating that the selected sample size is a cost-effective choice.

\begin{figure}[!t]
\centering
\begin{subfigure}[b]{\linewidth}
    \centering
    \includegraphics[width=\linewidth]{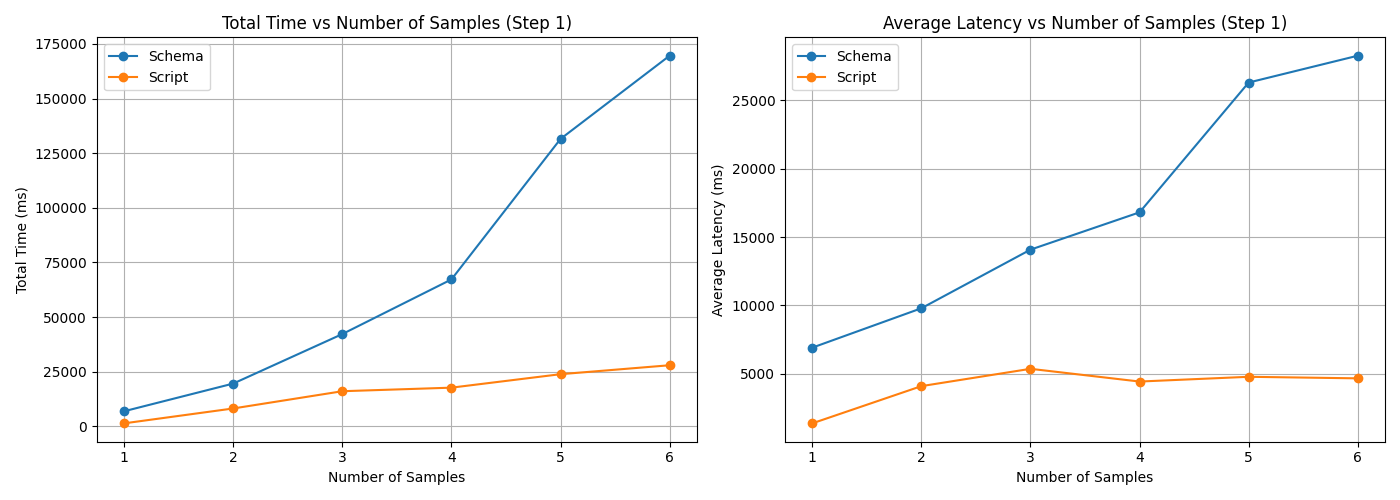}
    \caption{Logs}
    \label{fig:time_logs} 
\end{subfigure}

\begin{subfigure}[b]{\linewidth}
    \centering
    \includegraphics[width=\linewidth]{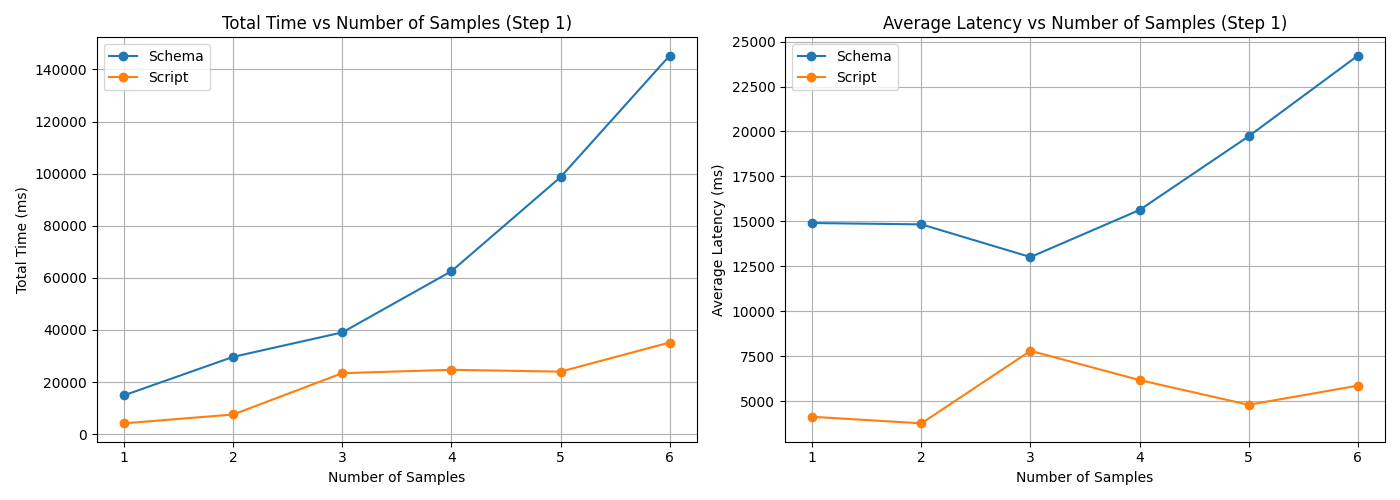}
    \caption{Configurations}
    \label{fig:time_configs} 
\end{subfigure}
\caption{Effect of samples size on prompt execution time (Step 1)}
\label{fig:step1_time}
\end{figure}

We also measured the prompt's average latency and total time, depicted in Figure \ref{fig:step1_time}. These results indicate that FastRAG processes the source data more quickly than submitting all data through the LLM. 
In particular, no prompt call required a retry for any sample size. Schema extraction typically takes longer than code generation, as schemas can be more complex and detailed. However, this process only occurs once at the beginning and remains unchanged unless significant changes, such as new services or applications, are introduced in the source data.
Similarly to character counts, the highest latency and total time values were associated with larger sample sizes, which are beyond the size of the samples we selected. Only a few requests required retries, mainly in cases where the schema was invalid due to errors like incorrect property types or missing required properties. 
An unusual spike in the average latency during script learning for the configuration dataset likely resulted from the LLM quickly producing an overly complicated schema that required more time or retries during script learning. This likely led to an imperfect parsing function, as indicated by the coverage ratio of $97\%$.

In general, when the average latency does not increase with additional sample chunks, it suggests that the added sample chunk did not introduce significant new information, allowing the LLM to quickly return an updated schema without much additional processing time.

\subsubsection{Step 2 script learning}
The process of determining the number of sample chunks for Step 2 script learning is similar to that of Step 1, but somewhat simpler. In Step 2, a parsing function is generated for each section, and no schema extraction is involved. As a result, the focus is solely on identifying the number of chunks needed to achieve a coverage ratio close to $100\%$. 
For simplicity, we fixed the same number of sample chunks for all sections, although the optimal number might vary between sections. However, the results demonstrate that the number of sample chunks does not need to be precisely accurate; achieving near $100\%$ coverage is possible even with one or two additional chunks beyond what is strictly necessary.

Table \ref{tab:step2_results} presents the performance metrics and parameters in Step 2 for both datasets. Note that the prompts generally required more retries than in Step 1. This increase in retries is due to the need to extract detailed properties for each section, which demands a higher level of precision and accuracy in the LLM-generated code.

\begin{table*}[h!]
\centering
\caption{Step 2 parameters and performance}
\label{tab:step2_results}
\begin{tabular}{|l|c|c|c|c|c|c|c|}
\hline
 & \textbf{Samples Size} & \textbf{Entity Types} & \textbf{Total Requests} & \textbf{Total In Chars} & \textbf{Total Out Chars} & \textbf{Total Time (s)} & \textbf{Coverage}\\ \hline
\textbf{Logs} & 4 & 7 & 33 & 153833 & 55763 & 182 & $98\%$ \\ \hline
\textbf{Configurations} & 2 & 9 & 20 & 56019 & 42945 & 111 & $100\%$ \\ \hline
\end{tabular}
\end{table*}

\subsection{Question-Answering}
We evaluate the retrieval effectiveness of FastRAG through question-answering. 
To generate relevant questions, we randomly selected evaluation chunks in Step 1 that were different from those used as sample chunks. We then used two different LLMs, namely GPT-4o-mini and Claude-3.5-Sonnet, to generate questions and answers based on these evaluation chunks. 
The questions were submitted to FastRAG, and we manually checked the accuracy of the answers. This manual verification is feasible given the relatively small size of the datasets and the manageable number of questions.

We submitted the questions to FastRAG with each of the four proposed retrieval methods —Graph, Text, Combined, and Hybrid.
We categorized each answer with one of the following labels:
(-) for incorrect answers (including hallucinations),
(+) for correct answers without any additional information, and
(++) for correct answers that include correct additional information. 

For the logs dataset (Table \ref{tab:qa_results_logs}), Graph querying retrieval struggled with questions requiring a broader contextual understanding, such as identifying types of logs or understanding the function of \textit{nova-compute.log} entries. This method often failed to retrieve comprehensive information, especially when abstract or holistic answers were needed. Text retrieval performed better in retrieving specific details directly from the data, such as typical entries in \textit{nova-api.log}. However, it fell short when exhaustive information was required, as seen in queries about common HTTP methods in the API logs. Combined and Hybrid retrieval outperformed Graph and Text methods by aggregating information from multiple sources, providing context-rich and detailed answers. Combined retrieval excelled, particularly in identifying the most common HTTP methods and providing a complete explanation of log functions. In some instances, guiding Text retrieval with query reformulations, such as broadening the search criteria, improves accuracy.

In the configurations dataset (Table \ref{tab:qa_results_configs}), Graph retrieval performed better, particularly in identifying devices and extracting specific configuration details such as IP addresses and route map values. This highlights its effectiveness in relational and accurate data contexts. However, it showed limitations in queries about access lists, possibly due to incomplete data extraction. Text retrieval, meanwhile, frequently produced incomplete results, particularly for queries such as identifying IP addresses or autonomous systems. Once again, Combined retrieval outperformed Graph and Text, providing more complete answers for queries like listing prefix lists, thanks to its ability to leverage both text search and graph-based methods. 
Hybrid retrieval, though effective in most cases, occasionally struggled with specific queries, such as access-list matches, pointing to a sensitivity in query formulation. 

Overall, combining graph-based and text-based retrieval provides the most effective solution for question-answering tasks in network data, allowing for both precision and completeness. However, while Hybrid retrieval currently underperforms compared to the Combined approach, it shows promising potential for improvement through techniques such as prompt refinement and few-shot examples.

\begin{table*}
\centering
\caption{Q\&A evaluation results for the logs dataset}
\begin{tabular}{|p{12cm}|c|c|c|c|}
\hline
\textbf{Question} & \textbf{Graph} & \textbf{Text} & \textbf{Combined} & \textbf{Hybrid} \\
\hline
What type of log APIs are present in the data & \cellcolor{green!30}+ & \cellcolor{red!30}- & \cellcolor{green!30}+ & \cellcolor{green!30}+ \\
\hline
What is the primary function of the nova-compute.log entries & \cellcolor{red!30}- & \cellcolor{green!30}+ & \cellcolor{orange!50}++ & \cellcolor{green!30}+ \\
\hline
What kind of information is typically logged in the nova-api.log entries & \cellcolor{red!30}- & \cellcolor{orange!50}++ & \cellcolor{green!30}+ & \cellcolor{green!30}+ \\
\hline
What are the most common HTTP methods observed in the api logs & \cellcolor{green!30}+ & \cellcolor{red!30}- & \cellcolor{orange!50}++ & \cellcolor{green!30}+ \\
\hline
What types of IP addresses are frequently seen in the log entries & \cellcolor{green!30}+ & \cellcolor{green!30}+ & \cellcolor{orange!50}++ & \cellcolor{green!30}+ \\
\hline
What kind of metadata-related requests are present in the logs & \cellcolor{red!30}- & \cellcolor{orange!50}++ & \cellcolor{orange!50}++ & \cellcolor{red!30}- \\
\hline
Are there any indications of VM state changes in the logs & \cellcolor{red!30}- & \cellcolor{green!30}+ & \cellcolor{green!30}+ & \cellcolor{orange!50}++ \\
\hline
What is the typical response time for API requests in the logs & \cellcolor{green!30}+ & \cellcolor{red!30}- & \cellcolor{green!30}+ & \cellcolor{green!30}+ \\
\hline
Are there any error status codes present in the log entries & \cellcolor{red!30}- & \cellcolor{green!30}+ & \cellcolor{green!30}+ & \cellcolor{red!30}- \\
\hline
Are there any recurring patterns in the server requests & \cellcolor{orange!50}++ & \cellcolor{red!30}- & \cellcolor{green!30}+ & \cellcolor{orange!50}++ \\
\hline
What is the latest status of the image with ID `0673dd71...` & \cellcolor{red!30}- & \cellcolor{red!30}- & \cellcolor{red!30}- & \cellcolor{red!30}- \\
\hline
How long the latest GET request to `/openstack/v2` takes & \cellcolor{red!30}- & \cellcolor{green!30}+ & \cellcolor{green!30}+ & \cellcolor{red!30}- \\
\hline
What events were logged for the instance with ID `af9d460c-89bf-...` & \cellcolor{red!30}- & \cellcolor{green!30}+ & \cellcolor{green!30}+ & \cellcolor{green!30}+ \\
\hline
How many times was the GET request to `/v2/54fadb41.../servers/detail` made & \cellcolor{green!30}+ & \cellcolor{red!30}- & \cellcolor{green!30}+ & \cellcolor{green!30}+ \\
\hline
What server IP addresses were involved in requests about metadata & \cellcolor{red!30}- & \cellcolor{red!30}- & \cellcolor{red!30}- & \cellcolor{red!30}- \\
\hline
What was the response length for the GET request to `/openstack/2013-10-17/meta\_data.json` & \cellcolor{green!30}+ & \cellcolor{green!30}+ & \cellcolor{green!30}+ & \cellcolor{green!30}+ \\
\hline
\end{tabular}
\label{tab:qa_results_logs}
\end{table*}

\begin{table*}
\centering
\caption{Q\&A evaluation results for the configurations dataset}
\begin{tabular}{|p{12cm}|c|c|c|c|}
\hline
\textbf{Question} & \textbf{Graph} & \textbf{Text} & \textbf{Combined} & \textbf{Hybrid} \\
\hline
How many devices are in the network & \cellcolor{green!30}+ & \cellcolor{red!30}- & \cellcolor{green!30}+ & \cellcolor{green!30}+ \\
\hline
What are the interfaces on `as1border1` & \cellcolor{green!30}+ & \cellcolor{red!30}- & \cellcolor{green!30}+ & \cellcolor{green!30}+ \\
\hline
What is the IP address of interface `GigabitEthernet0/0` on `as1border1` & \cellcolor{green!30}+ & \cellcolor{red!30}- & \cellcolor{green!30}+ & \cellcolor{red!30}- \\
\hline
What is the typical local-preference value set in the route maps & \cellcolor{green!30}+ & \cellcolor{green!30}+ & \cellcolor{green!30}+ & \cellcolor{green!30}+ \\
\hline
What is the local preference value set in the route-map `as2\_to\_as1` & \cellcolor{green!30}+ & \cellcolor{red!30}- & \cellcolor{green!30}+ & \cellcolor{green!30}+ \\
\hline
Which access-list permits IP traffic for the host `1.0.2.0` & \cellcolor{red!30}- & \cellcolor{green!30}+ & \cellcolor{green!30}+ & \cellcolor{red!30}- \\
\hline
What prefix-list is matched in the route-map `as2\_to\_as3` & \cellcolor{green!30}+ & \cellcolor{red!30}- & \cellcolor{green!30}+ & \cellcolor{red!30}- \\
\hline
What are the prefix-lists configured & \cellcolor{green!30}+ & \cellcolor{green!30}+ & \cellcolor{orange!50}++ & \cellcolor{green!30}+ \\
\hline
\end{tabular}
\label{tab:qa_results_configs}
\end{table*}

\subsection{Cost and Speed Improvement}
We compared the time and cost of FastRAG with those of GraphRAG discussed in Section \ref{rw}, through the processing of logs and configuration data.

To maintain a manageable experiment size while still having enough source data for comparison, we adapted the original datasets as follows: 
the logs dataset was reduced from 2000 to 500 lines, resulting in 78 chunks (of $\sim1000$ tokens) for GraphRAG and 319 chunks (of $\sim1000$ tokens) for FastRAG (Step 1). 
The configuration dataset was replaced with a larger dataset \cite{netplumber}, comprising five configuration files totaling 4,400 lines, which yielded 67 chunks of $\sim1000$ tokens for both GraphRAG and FastRAG (Step 1).

To ensure similar conditions for comparison, we adjusted the parameters of GraphRAG as follows: the chunk size was set to 1000 tokens, the chunk overlap was set to 50 tokens ($\sim5$ lines), and the number of concurrent requests was limited to one \footnote{GraphRAG improves latency by supporting parallel requests to the LLM, an engineering feature that FastRAG could also support but is not implemented in the current prototype}.

We measured the total time and cost\footnote{Cost measured from the API usage provided on the OpenAI dashboard.} required to run both RAG systems until the KG creation.
For FastRAG, the measured time included the $(n_c,n_t)$ hyperparameter search for the fixed number of sample chunks for each dataset. The procedure of determining the optimal sample size through trial and coverage was not included in the total time.
The best sample chunk sizes were fixed as follows. 
For the configurations dataset: 6 sample chunks in Step 1 (achieving $98.7\%$ coverage) and 4 sample chunks in Step 2 (achieving $94.6\%$ coverage). 
For the logs dataset: 4 sample chunks in Step 1 (achieving $100\%$ coverage) and 4 sample chunks in Step 2 (achieving $95\%$ coverage).

Table \ref{tab:improvement} presents the results for both evaluated systems. As expected, the difference in cost and time between FastRAG and GraphRAG is significant, even with these relatively small datasets. FastRAG's improvement is evident, particularly when considering that the generated parsing functions do not need to be regenerated, as long as the source data contains the same type of information. These functions could also be periodically updated to refine the information extraction.

Note that GraphRAG uses LLMs to generate entity descriptions and community reports, which are essential for question-answering.
Additionally, while we limited GraphRAG to one concurrent LLM request for fairness, in practical scenarios, the processing time could be reduced by using parallel requests. However, this does not reduce the cost.

\begin{table}[h!]
\centering
\caption{Comparison of FastRAG and GraphRAG}
\begin{tabular}{|c|c|c|c|c|}
\hline
\textbf{Dataset} & \textbf{System} & \textbf{Time (min)} & \textbf{Cost (USD)} \\ \hline
\multirow{2}{*}{Logs} & FastRAG & 3.9 & 0.91 \\ \cline{2-4} 
                           & GraphRAG & 40.0  & 6.00 \\ \hline
\multirow{2}{*}{Configurations} & FastRAG & 2.0 & 1.10 \\ \cline{2-4} 
                           & GraphRAG & 6.3 & 4.00 \\ \hline
\end{tabular}
\label{tab:improvement}
\end{table}

\section{Conclusion}\label{conclusion}
FastRAG is a cost-effective and time-efficient method for processing semi-structured network data, demonstrated on network logs and configurations. Its reliance on sample selection introduces the risk of missing small portions of information ($< 5\%$ over the reported results). Furthermore, as an LLM-powered system, it introduces variability, which requires continuous verification to ensure accuracy. The current design also lacks extraction of entity relationships, which could improve information retrieval at a slightly higher cost. 

Despite its limitations, FastRAG offers promising performance and resource savings for processing large and frequently changing data, particularly given the rapidly improving LLM capabilities. At the time this study was completed, OpenAI announced a new version of GPT-4o, which reliably adheres to developer-supplied JSON schemas and a new series of models designed to spend more time reasoning before generating their responses.

\section*{Disclaimer}
Any mention of commercial products or references to commercial organizations is for information only; it does not imply NIST recommendation or endorsement, nor does it imply that the products mentioned are necessarily the best available for that purpose.

\bibliographystyle{IEEEtran}
\bibliography{biblio.bib}

\end{document}